\documentclass[prld,showpacs,showkeys,twocolumn,notitlepage,10pt]{revtex4-1}
\usepackage{amsfonts,bbold,amsmath,amssymb,graphicx,epstopdf,verbatim,dsfont,color}
\usepackage[english]{babel}

\def \be{\begin{equation*}}
\def \ee{\end{equation*}}

\begin{document}

\title{Can math beat gamers in Quantum Moves?}
\author{Dries Sels}
\affiliation{Department of Physics, Boston University, Boston, MA 02215, USA}

\affiliation{Department of Physics, Harvard University, 17 Oxford st., Cambridge, MA 02138, USA}
\affiliation{Theory of quantum and complex systems, Universiteit Antwerpen, B-2610 Antwerpen, Belgium}
\date{\today}

\begin{abstract}
In a recent work on quantum state preparation, S{\o}rensen and colleagues\cite{sorensen_16} explore the possibility of using video games to help design quantum control protocols. The authors present a game called "Quantum Moves"\cite{quantummoves_16} in which gamers have to move an atom from A to B by means of optical tweezers. They report that, \emph{players succeed where purely numerical optimization fails}\cite{sorensen_16}. Moreover, by harnessing the player strategies they can \emph{outperform the most prominent established numerical methods}\cite{sorensen_16}.
The aim of this manuscript is to analyze the problem in detail and show that those claims are untenable. In fact a simple stochastic local optimization method can easily find very good solutions to this problem in a few 1000 trials rather than the astronomical $7.4\times 10^{8}$ trials of the most successful optimization method reported in \cite{sorensen_16}.  Next, counter-diabatic driving is used to generate protocols without resorting to numeric optimization\footnote{As E. Wigner once said:  \emph{It is nice to know that the computer understands the problem. But I would like to understand it too.}}; the protocols are shown to outperform virtually all players.  The analysis moreover results in an accurate analytic estimate of the quantum speed limit which, apart from zero-point motion, is shown to be entirely classical in nature. The latter might explain why gamers are remarkably good at the game.
\end{abstract}

\maketitle

\section*{Introduction}
\begin{figure}
\includegraphics[width=0.9\columnwidth]{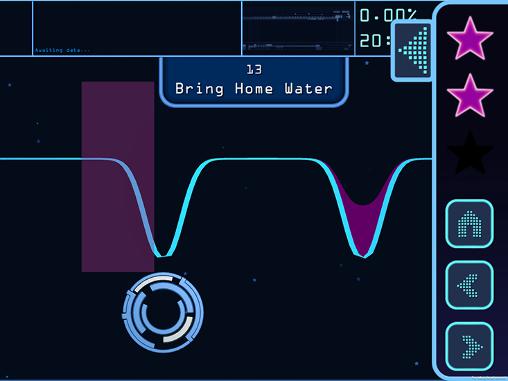}
\caption{{\bf Quantum moves interface} Screenshot of the BringHomeWater challenge developed by S{\o}rensen et al.~\cite{sorensen_16}. In the game players move an optical tweezer (blue circle is the cursus to move the tweezer) from an initial point on the left towards an atom trapped in a potential well on the right. The atom is depicted as the purple liquid filling up the potential-well. Players have to find a route to bring this liquid to the target (purple shaded region on the left) as fast as possible without spilling it.}
\label{fig:moves}
\end{figure}

The entire gamification research program---the use of games to channel human brain-power to solve research problems---is based on the sole premise that \emph{"Humans are better than computers at certain tasks because of their intuition and superior visual processing"}~\cite{maniscalco_16}. Previous examples of citizen-science games such as Foldit~\cite{foldit_10} and Eyewire~\cite{eyewire_14} are 3-D puzzle games, think of them as digital Rubik's cubes.  
Their success can indeed be attributed to our superior visual insight in 3D shapes. But what if our visual processing is impaired by our inability to imaging anything in more than 3 dimensions? Or our intuition about the problem is simply wrong. Are our heuristics any better than machines at that point? 
Recent progress in machine learning seems to indicate they are not. With almost super-human performance in Go~\cite{alphago_16}, computers have mastered to most complex of classical games using a combination of reinforcement learning and supervised learning. Deep convolutional neural networks have resulted in unseen machine performance in the visual domain~\cite{hinten_12} and deep reinforcement learning~\cite{atari_15} has produced a computer agent which rivals professional players in many Atari 2600 games.  
In light of these results one might wonder how players of Quantum Moves (Fig.~\ref{fig:moves}) were able to outperform purely numerical optimization methods designed for quantum optimal control problems. Given the bizarre and counterintuative nature of quantum mechanics one could even wonder how players came up with a reasonable strategy to begin with. 

The goal of this work is to elucidate this tension. The manuscript starts by introducing the concept of counter-diabatic driving~\cite{sels_17}. The latter formalizes the intuitive notion of players to move as fast as possible without spilling~\cite{maniscalco_16}. The idea can readily be applied to a single optical tweezer and it is shown how this completely predicts the quantum speed limit. A recent result by Jarzynski et al.~\cite{jarzynski_17} is used to generalize it to the double tweezer situation. Protocols are compared to those obtained from purely numeric optimization and the convergence of the numeric algorithm is discussed. Finally, this manuscript describes how tunneling is not a viable strategy.

\section*{Quantum state preparation}
\begin{figure}
\includegraphics[width=\columnwidth]{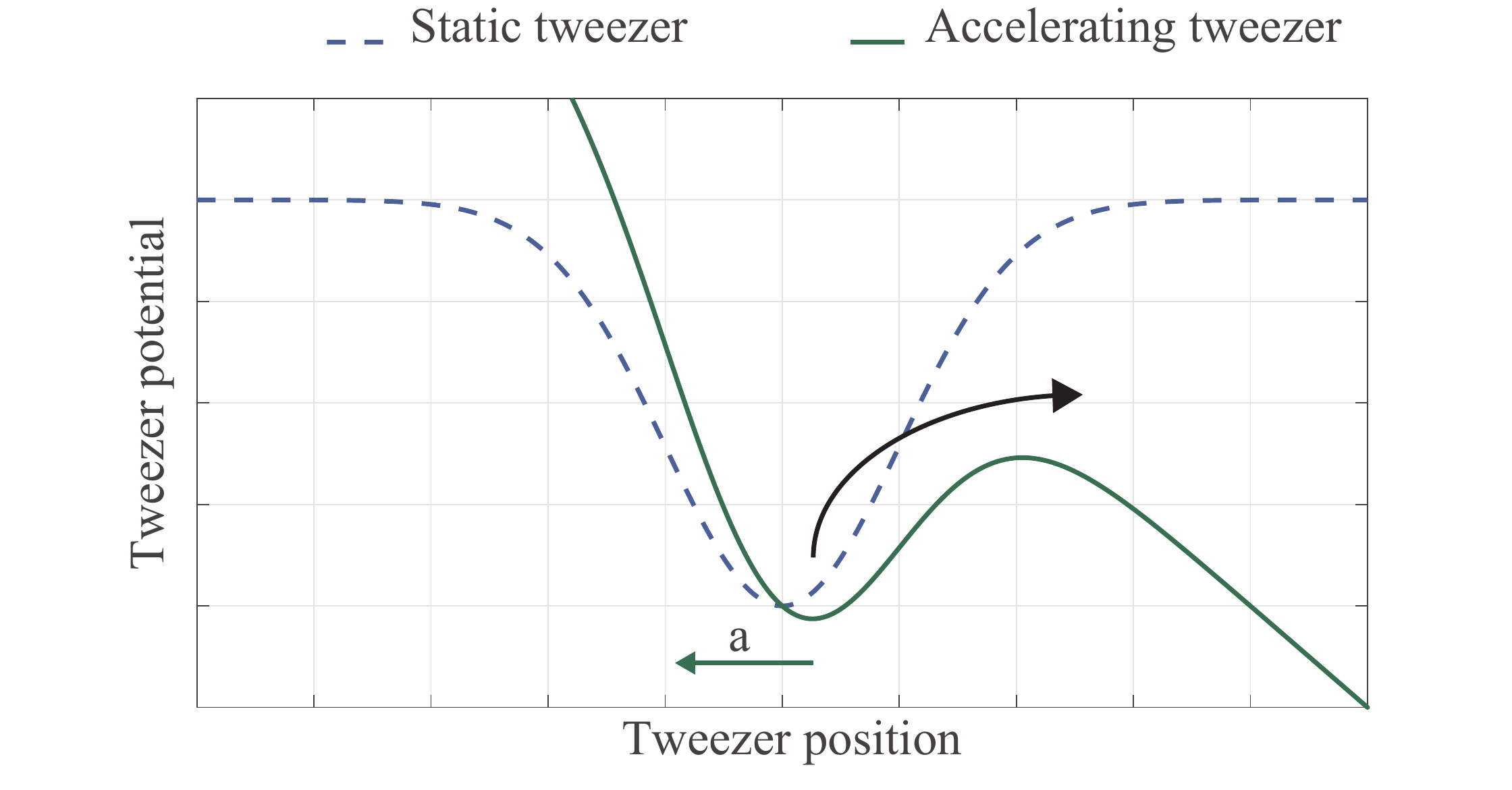}
\caption{{\bf Effective potential for moving objects} The potential of a static optical tweezer is depicted by the dashed blue line. The atom however feels a different potential once the tweezer is moving. By going to the co-moving frame the atom picks up an additional pseudo-force if the tweezer accelerates, just like a person in an elevator. For a tweezer of finite depth, it unfortunately allows atoms to escape if the acceleration is too large.}
\label{fig:potential}
\end{figure}

Let's consider first what happens to an atom in the ground state of just a single optical tweezer, i.e. the system's Hamiltonian is
\be
H=\frac{p^2}{2m}+V(x-x_0(t)),
\ee
where $V(x)$ denotes the tweezer potential (Fig.~\ref{fig:potential}) and $x_0(t)$ the tweezer's position. The goal is to move the tweezer without exciting the system. Since the entire system is simply being translated in space, the adiabatic gauge potential~\cite{mike_17} is the momentum operator, i.e. the generator of translations. Consequently, the counter-diabatic Hamiltonian becomes
\be
H_{\rm CD}=\frac{p^2}{2m}+V(x-x_0(t))+\dot{x_0}p
\ee
For a more detailed discussion on counter-diabatic driving I refer to ref.~\cite{sels_17}. Note that we can gauge transform this into an equivalent Hamiltonian which accomplishes the same task as long as we demand that the velocity $\dot{x_0}$ vanishes in the beginning and at the end of the protocol. Just like in electromagnetism, we turn the curl-free vector potential into a scalar potential, which results in
\begin{equation}
H'_{\rm CD}=\frac{p^2}{2m}+V(x-x_0(t))-m\ddot{x}_0x
\label{eq:H_CD}
\end{equation}
As expected the atom can be kept in the ground state even when the tweezer is moving as long as we cancel out the \emph{pseudo-force} due to the acceleration of the system (Fig.~\ref{fig:potential}). This is not completely within our reach since we can only tune the depth and the position of the tweezer. The best we can do is, shift the tweezer a bit such that it effectively generates the counter-diabatic force on the atom. Since the trap is not infinitely deep, the maximal force it can generate is limited and counter-diabatic driving will fail once the pseudoforce exceeds this maximal value. Counter-diabatic driving can only work as long as
\be 
\max_x | \partial_x V(x)|>ma,
\ee
where a is the maximal acceleration. For a Gaussian beam with amplitude $\mathcal{A}$, and standard deviation $\sigma$ 
\be
a<\frac{\mathcal{A}}{m\sigma\sqrt{e}}.
\ee
The smaller the acceleration, the better. One way to do so is of course to go to the adiabatic limit but in this case we want to do it in fixed duration $T$. We thus want a protocol $x_0(t)$ that covers a distance $L$, let's say from $x_i=L/2$ to $x_f=-L/2$, with the smallest magnitude of the acceleration. This would of course be a straight line or geodesic. Recall however, that the velocity had to vanish at the beginning and the end of the protocol as well. A simple protocol that accomplishes all of the above requirements is a cubic polynomial:
\be
x_0(t)=L \left(2 \left(\frac{t}{T}\right)^3-3\left(\frac{t}{T}\right)^2+\frac{1}{2}\right).
\ee
This protocol has maximal acceleration at the beginning and the end, namely $a=6L/T^2$. Combining this with the above restriction on the acceleration we find a constraint on the protocol duration,
\be
\boxed{T>T_{\rm CSL}=\left(\frac{6mL\sigma\sqrt{e}}{\mathcal{A}} \right)^{1/2}}
\ee
Plugging in the actual parameters used in Ref.~\cite{sorensen_16}, being $\mathcal{A}=130-160$, $m=1$, $\sigma=1/8$, $L=1.1$, we find $T_{\rm CSL}=0.092-0.102$. In perfect agreement with the numeric results of~\cite{sorensen_16}. To find the approximate counter diabatic protocol one can simply approximate the tweezer potential by a harmonic potential with frequency  $m\omega^2=\mathcal{A}/\sigma^2$. For the harmonic oscillator the counter-diabatic protocol is simply given by~\cite{mike_17}
\begin{equation}
x_{\rm CD}(t)=x_0(t)+\frac{\ddot{x}_0(t)}{\omega^2}.
\label{eq:single_CD}
\end{equation}
Moreover, by eliminating the tweezer amplitude in favor of the effective trap frequency, the classical speed limit becomes
\be
T_{\rm CSL}\approx \frac{\pi}{\omega} \left( \frac{L}{\sigma}\right)^{1/2}
\ee

\begin{figure}
\includegraphics[width=\columnwidth]{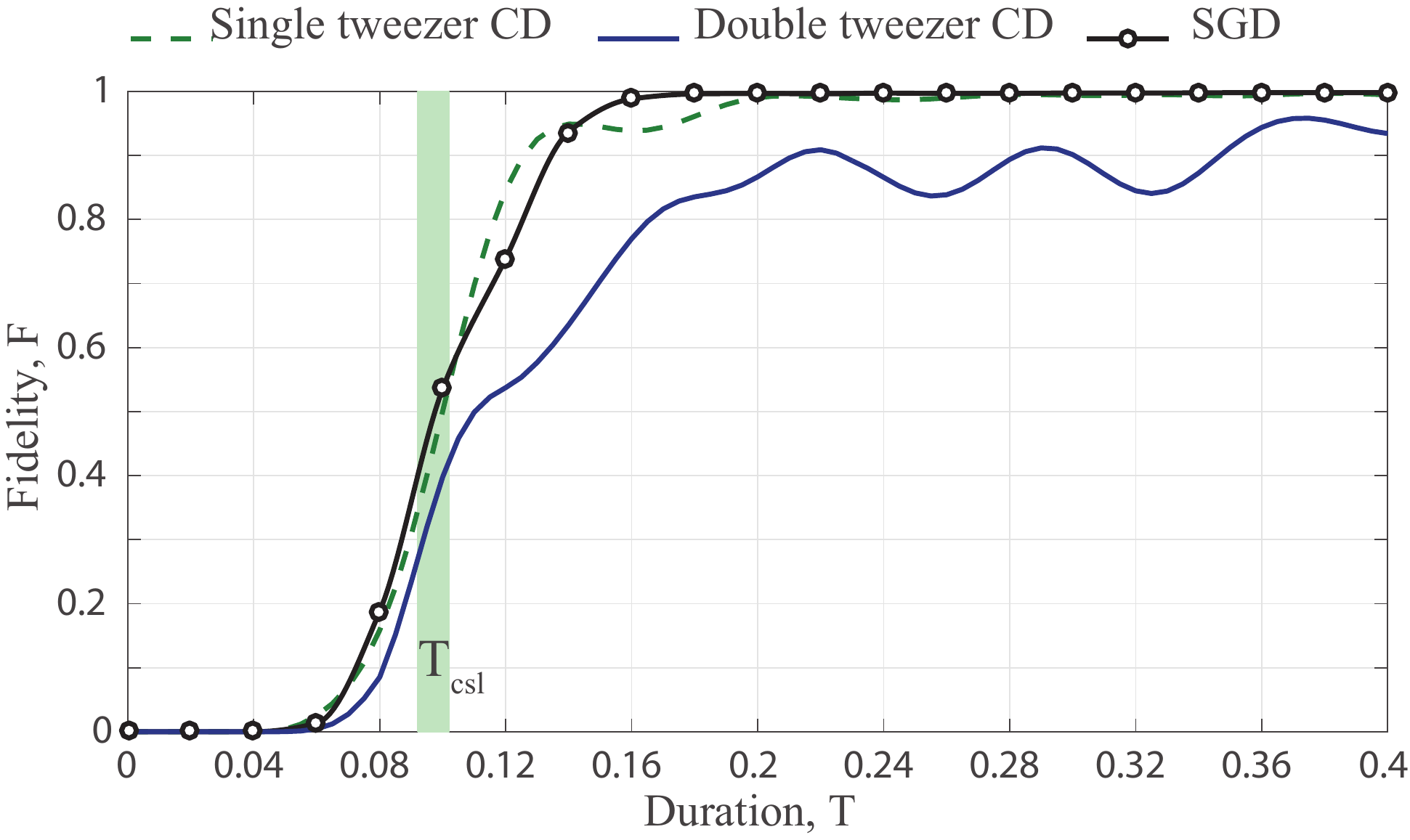}
\caption{{\bf Fidelity for different protocol durations} The green shaded region indicates the classical speed limit for moving a particle in a finite depth potential. The fidelity of the associated quantum counter-diabatic driving (dashed green) protocol shows a sharp rise of fidelity near $T_{\rm CSL}$, confirming that zero-point motion is the only relevant quantum effect in the problem. For the double tweezer the protocol is adapted to account for an additional geometric contribution to the gauge potential. Up to $T_{CSL}$, its performance (blue line) is comparable to that obtained from numeric optimization (black dots). At later times, its performance starts to wane because it is derived from an approximate adiabatic gauge potential. Details on the numeric optimization algorithm can be found in the Methods section.}
\label{fig:Fid}
\end{figure}

Up to this point the derivation was completely classical, which has resulted in the complete absence of Planck's constant in the minimal required duration. One might wonder, how quantum mechanics will change these results. If we postpone the possibility of tunneling between the two tweezers for a second, and simply focus on a single moving tweezer, the quantum mechanical effects would be small. A simple argument goes as follows, for the harmonic oscillator there are no dynamical quantum effects. Its dynamics is completely classical and quantum mechanics only puts constraints on the allowed initial states. 
On the timescale $T_{CSL}$ a particle can only go around the trap about 1-2 times so any quantum interference due to the anharmonicity of the trap is small. It's therefore reasonable to say that the only relevant quantum effect is the zero-point fluctuations in the initial state. Instead of having a sharp transition from perfectly controllable to completely uncontrollable there will be a smooth crossover around $T_{CSL}$. Numerical simulation of single tweezer counter-diabatic result confirms this argument (Fig.~\ref{fig:Fid}).

\begin{figure}
\includegraphics[width=\columnwidth]{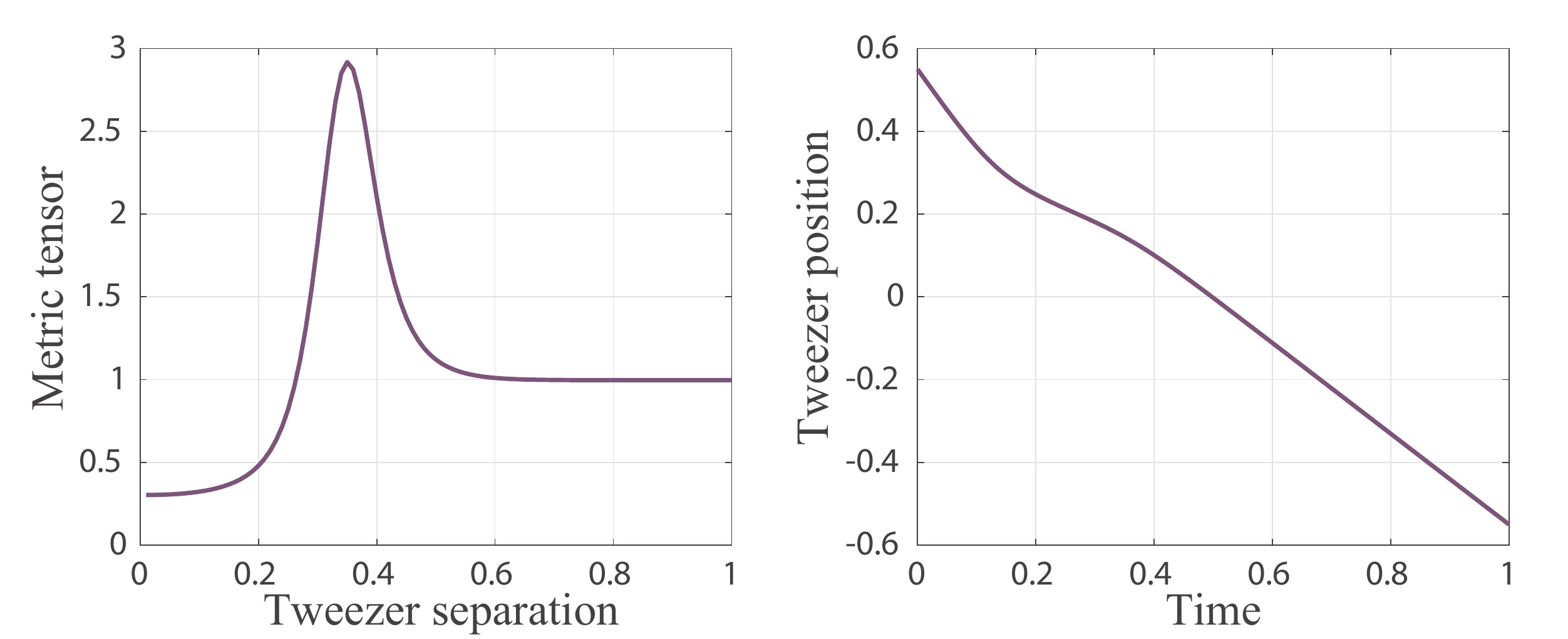}
\caption{{\bf Geometry and geodesic protocol} The left panel shows the effective adiabatic metric $g$ for moving the tweezer. In the middle there is a region where a small change in the position of a tweezer can cause a large change in the shape of the potential energy, at which point the metric becomes much larger than one. The resulting geodesic is shown in right panel; note how it slows down where the metric is large.}
\label{fig:metric}
\end{figure}

So far we have only been concerned with a single tweezer. While numerical results indicate that the fidelity for the single tweezer is comparable to that of the actual two tweezer problem, it remains a challenge to find the actual protocol. In the two tweezer problem, the original tweezer in which the atom is trapped can not be moved, and a second movable tweezer has to transport the atom (Fig.~\ref{fig:moves}). It would be unwise the simply use~\eqref{eq:single_CD} as it hinges on the fact that the adiabatic gauge potential is simply the momentum operator. The adiabatic gauge potential for the two tweezer problem is much more complicated and in general there is no hope that we can turn it into an effective scalar potential like before. For the ground-state of the system, it is however possible to design an approximate counter-diabatic gauge field that captures almost all of the physics and constitutes only a small (but important) correction to single tweezer result:
\begin{center}
\begin{tabular}{ c c}
single tweezer &$\Leftrightarrow$ double tweezer \\
$\dot{x}_0 p$ & $\dot{x}_0 \sqrt{g(x_0)} p$ 
\end{tabular}
\end{center} 
The exact details are given in section Methods, but the result has a very intuitive explanation. The additional correction $\sqrt{g}$ is caused by the fact that the equilibrium position of an atom in the double tweezer potential does not always coincide with the minimum of the potential of the moving tweezer. In other words $g$ serves as an effective metric tensor, which contracts and dilates space to account for the discrepancy between tweezer's and the atom's position. Since the gauge potential is the generator of adiabatic transformations, it translates the atom to its new equilibrium position as the tweezer is moved. When the two tweezers are close together the equilibrium position shifts slower than the tweezer, such that $g<1$. For two far separated tweezers the single tweezer result should hold, implying $g\rightarrow1$. In the middle, however, there is a region where the minimum of the potential quickly shifts between the two tweezers. The latter results in a significant peak in the metric (Fig.~\ref{fig:metric}).
One can thus avoid most non-adiabatic effects by moving the tweezer on a geodesic of the metric $g$. Just like for the single tweezer one has to make sure that the velocity goes to zero in the beginning and the end of the protocol. At present, this is done by simply adding a region of constant acceleration at the beginning and the end. Just like for the single tweezer we find an effective protocol by correcting the trial protocol for its deviation from the geodesic:
\begin{equation}
x_{\rm CD}(t)=x_0(t)+\frac{1}{\omega^2} \frac{\partial }{\partial t} \left( \sqrt{g(x_0)}\frac{\partial x_0(t) }{\partial t}  \right).
\label{eq:single_CD}
\end{equation}
The performance of these protocols is summarized in Fig.~\ref{fig:Fid}. Again there is a sharp rise of the fidelity near the classical speed limit. For longer protocol durations, it's performance starts to falter. This suboptimal performance is entirely caused by the approximate nature of the adiabatic gauge potential. 

\begin{figure}
\includegraphics[width=\columnwidth]{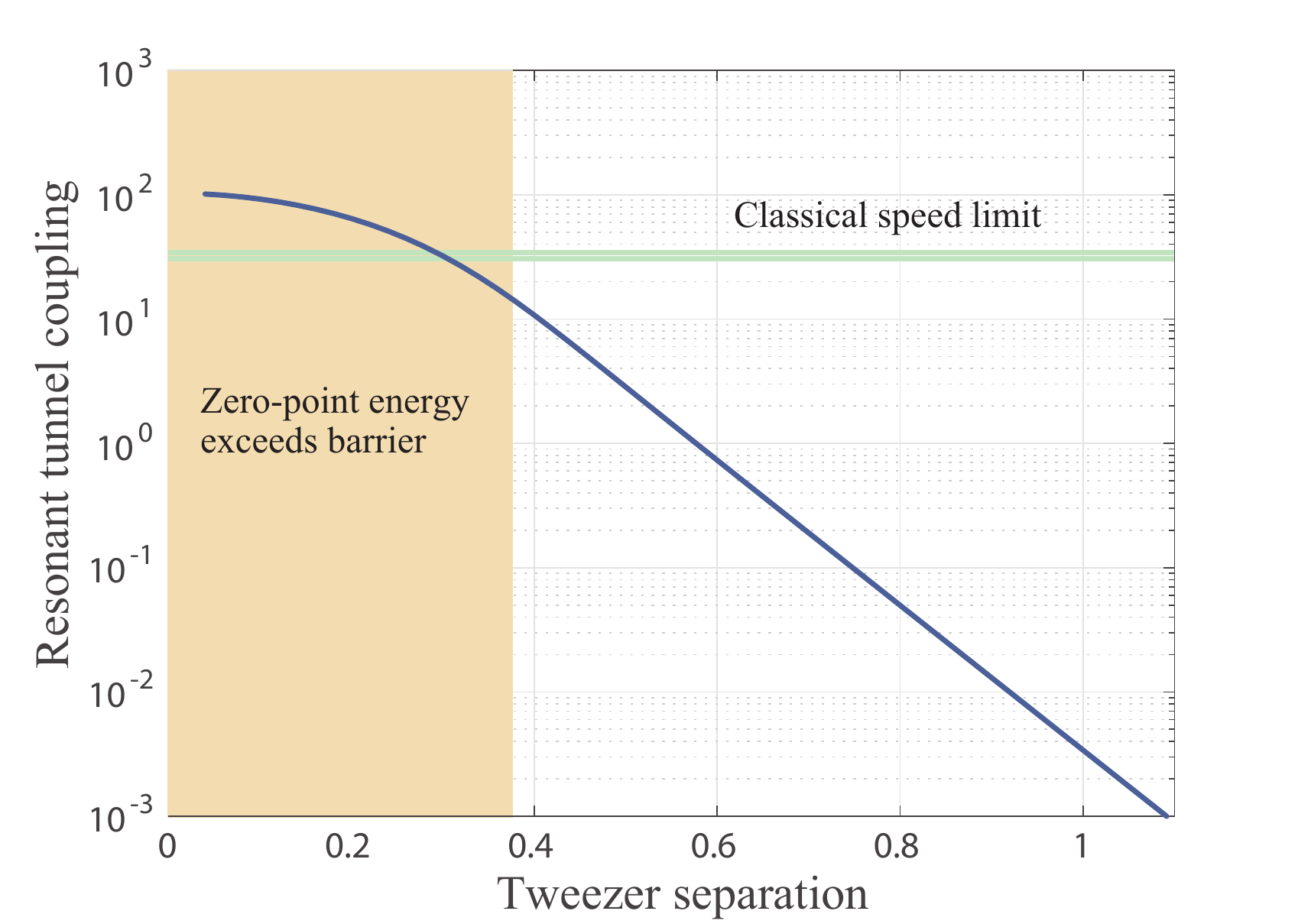}
\caption{{\bf Tunnel coupling} The blue line shows the tunnel matrix element between the ground states of the two tweezers. As a reference, the coupling which transfers the population in the classical speed limit, i.e. $E=\pi/(T_{CSL})$, is shown (green shaded region).  At sufficiently large separation the tunnel coupling decays exponentially with $\kappa=\sqrt{-2mE_0}/\hbar$, where $E_0$ is the ground state energy.}
\label{fig:tunnel}
\end{figure}

Given these limitations we should investigate the possibility of direct tunneling between the tweezers. Unlike the moving tweezer, this approach is explicitly non adiabatic and it is not limited by excitations caused by the pseudo-force. On the other hand the tunnel coupling between the two tweezers is exponentially suppressed in there separation. At sufficiently large distance it can therefore never outperform the moving tweezer since the classical speed limit only scales like $L^{1/2}$. For the present parameters, the (resonant) tunneling coupling as a function of tweezer separation is shown in Fig.~\ref{fig:tunnel}. In a time $T_{CSL}$ it is only possible to transfer the population over a distance of $0.3$, that is not nearly the required $1.1$ and is even in the range where the ground state energy exceeds the potential barrier height between the two tweezers. Although interference plays an important role in this regime, there is no tunneling. At present tunneling only happens when $L\gtrsim 3\sigma$, which already takes $T=0.2$ to complete the state transfer. It's therefore save to conclude that Quantum Moves gamers should not devise strategies that \emph{go beyond the classical laws of physics}~\cite{maniscalco_16}. This explains why gamers are quite good at the game, since the best strategy is indeed the intuitive one where you try to move the system without spilling the water.

\section*{Methods}
\subsection*{State specific counter-diabatic driving}
Counter-diabatic driving is a method to speed up quantum state preparation for adiabatically connected states by adding an additional driving field to the Hamiltonian. In general, the additional Hamiltonian is hard to find. For chaotic systems, the exact field is even ill defined~\cite{jarzynski_95,mike_17}. Recently, some progress has been made to find approximate counter-diabatic drives for complex systems~\cite{sels_17}. At present, the situation is slightly simpler since we simply deal with ground states of a 1-D single particle problem. A recent idea by Jarzynski and colleagues~\cite{jarzynski_17} is particularly suited for this problem. 
The idea is very simple, i.e. consider a particle described by the wave function $\psi(x,t)$ and subject to the Hamiltonian:
\begin{equation*}
H_1(t)=\frac{1}{2} \left( v(x,t)p+pv(x,t) \right).
\end{equation*}
From the Schr\"odinger equation one readily derives the continuity equation for the density $n(x,t)=| \psi(x,t)|^2$,
\begin{equation}
\partial_t n(x,t)=-\partial_x \left(v(x,t) n(x,t) \right).
\label{eq:continue}
\end{equation}
Given a target time-dependent density, e.g. an adiabatic ground state, we are trying to find the velocity-field $v(x, t)$ such that we always recover this density. Instead of solving the continuity equation for the density we thus solve for the velocity-field, which results in
\begin{equation}
v(x,t)=-\frac{1}{n(x,t)} \int_{-\infty}^x {\rm d}\xi \partial_t n(\xi,t)=-\frac{\partial_t I(x,t)}{\partial_x I(x,t)},
\end{equation}
where $I(x,t)$ is the cumulative distribution function (cf. expression (12) in~\cite{jarzynski_17}). If the density is that of an eigenstate of a time dependent Hamiltonian $H_0(t)$ one can of course add the bare Hamiltonian to $H_1(t)$ to get the full counter-diabatic drive $H_{CD}=H_0+H_1$. While this will result in perfect state transfer, it will require access to arbitrarily complicated potential energies. In the end, we can only shift the tweezers and so we would like to restrict to spatially homogenous $v(x,t)=v(t)$, as they will result in a linear potential. To get the approximate velocity field one can simply solve eq.~\eqref{eq:continue} in the least square sense, i.e.
\begin{equation}
v(t)= \underset{w}{\rm argmin} \int {\rm d}x \left(\partial_t n+w\partial_x n \right)^2,
\end{equation}
which yields
\begin{eqnarray}
v(t)=-\frac{\int {\rm d}x (\partial_t n)(\partial_x n) }{ \int {\rm d}x \left( \partial_x n\right)^2}&=&\dot{x}_0 \frac{\int {\rm d}x (-\partial_{x_0} n)(\partial_x n) }{ \int {\rm d}x \left( \partial_x n\right)^2} \nonumber \\
&\equiv& \dot{x}_0 \sqrt{g(x_0)} \nonumber,
\end{eqnarray}
where we have used the fact that the ground-state density is fully parametrized by the position of the tweezer $x_0$. This moreover results in a robust numerical scheme to extract $v(t)$. Note that, for a single tweezer $\partial_{x_0}n=-\partial_x n$ such that $v(t)=\dot{x}_0$. For the latter the velocity field is exact, for any other situation it is only an approximation. Finally, one readily identifies $g$ with an effective adiabatic metric tensor, which dilates and contracts space to account for the shift between the tweezer position and the equilibrium position of the atom. Indeed, no pseudo-force will be generated whenever $v(t)$ is constant, a condition which exactly yields the geodesic equation:
\be
\ddot{x}_0+\frac{1}{2} \left( g^{-1}\partial_{x_0}g\right) \dot{x}_0^2=0.
\ee
 The metric tensor and geodesic for the present --- two tweezer problem --- can be found in Fig.~\ref{fig:metric}. The density is computed from the ground state of the Hamiltonian:
\begin{equation*}
 H(x_0)=\frac{p^2}{2}-\mathcal{A} \exp{\left(-\frac{(x-x_0)^2}{2\sigma^2}\right)}-\mathcal{B} \exp{\left(-\frac{x^2}{2\sigma^2}\right)},
\end{equation*}
 where $\mathcal{A}=160$, $\mathcal{B}=130$ and  $\sigma=1/8$, in agreement with the parameters reported in~\cite{sorensen_16}.
\subsection*{Stochastic ascend optimizer}
In order to store and optimize the protocols we have to reduce them to a finite number of degrees of freedom. This is done by considering quasi-continuous protocols of duration $\Delta t=T/N$. Moreover, we also quantize the allowed values of the tweezer position with a spacing $\Delta x=\sigma/8$ and allow it to move in the interval $[-1,1]$. This results in $M=128$ different tweezer positions $x^k$. For each of those positions we have a Hamiltonian $H_k$ and an associated unitary
\begin{equation}
U_k=\exp \left(-i H_k \Delta t \right).
\end{equation}
All those unitaries can be precomputed and stored, this only has to be done once. The total time evolution operator $U(T)$ for a protocol thus consists of a multiplication of a subset of these unitaries
\begin{equation}
U(T)=U_{k_N} U_{k_{N-1}} \ldots U_{k_2} U_{k_1}=\prod_{i=1}^N U_{k_i},
\end{equation}
and the fidelity for reaching the target state $\left\langle \phi \right |$ from the initial state $\left| \psi \right\rangle$ is given by $F=|\left\langle \phi \right| U \left| \psi \right\rangle|^2$. To optimize the fidelity we now simply do local stochastic ascend optimization, which works as follows: 
\begin{enumerate}
\item[0] Draw a random starting protocol $x=(x_N,x_{N-1},\ldots,x_2,x_1)$
\item[1] Randomly permute all integers $1:N$, let's call this vector $I$. This will ensure we run over all the timesteps in a random fashion but sweep over the entire protocol before we change a local setting again. Consider the current integer to be $W$
\item[2] Compute $\left\langle \phi_{W+1} \right|=\left\langle \phi \right| \prod_{i=W+1}^N U_{k_i}$ and $ \left| \psi_{W-1} \right\rangle= \prod_{i=1}^{W-1} U_{k_i} \left| \psi \right\rangle$
\item[3] Compute the fidelity $F_k=|\left\langle \phi_{W+1} \right| U_{k} \left| \psi_{W-1} \right\rangle|^2$ for all $k$.
\item[4] Select the optimal $k$ and replace the corresponding element in $x$ with $x^k$, i.e. $x_W=x^k$
\item[5] Repeat step 2,3,4 for all the elements  of $I$. If done, we have swept over the entire protocol once and forced the protocol to the best local settings. 
\item[6] Sweep over the protocol again by going back to step 1. Stop when tollerance is reached or when no element was changed in the previous sweep. The latter implies a local optimum is reached.
\end{enumerate}

\begin{figure}
\includegraphics[width=\columnwidth]{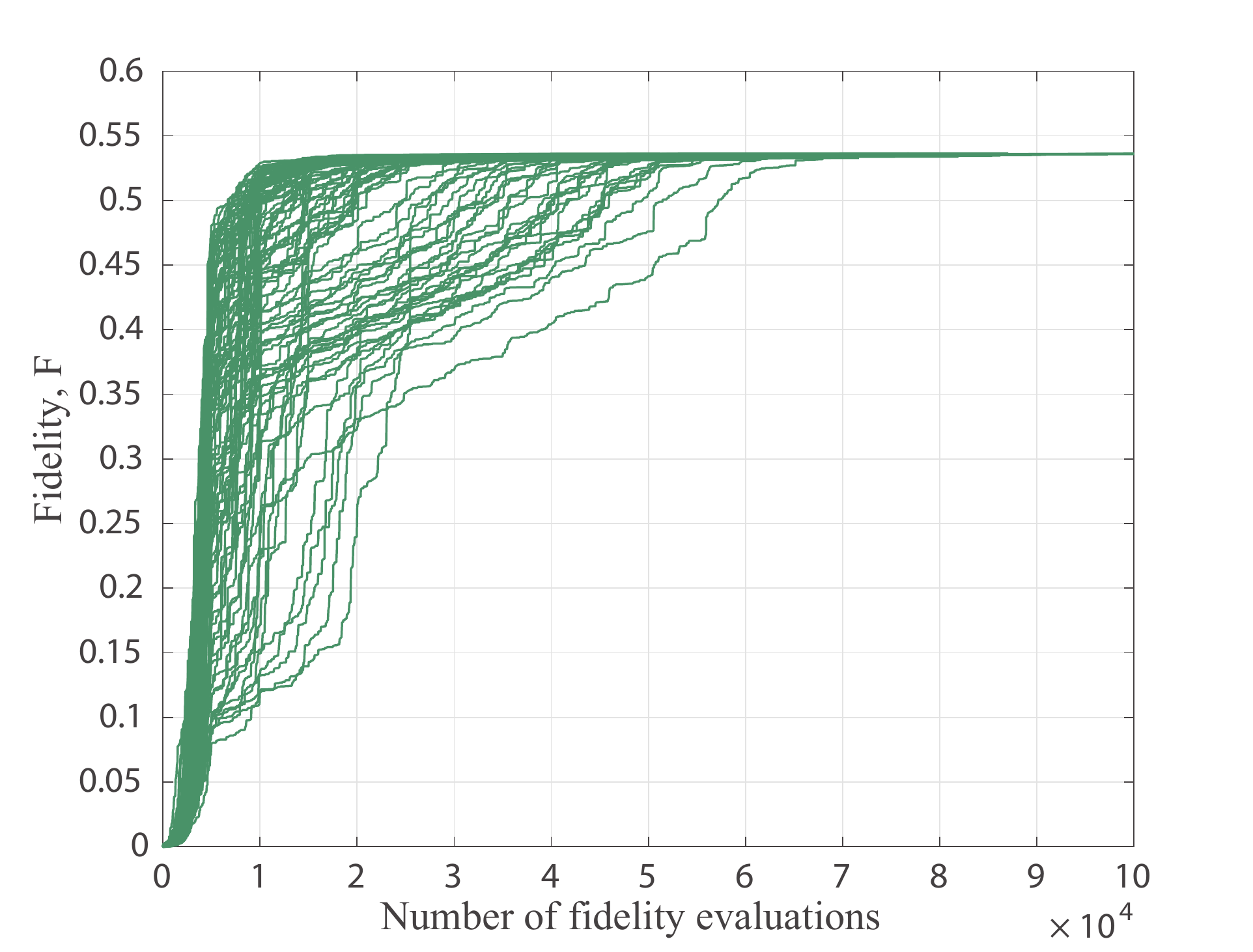}
\caption{{\bf Stochastic fidelity traces for T=0.1} Figure shows how 100 initial random protocols move through the fidelity landscape. There are $M=128$ settings and $N=40$ timesteps, such that a single sweep over the lattice results in $5120$ fidelity evaluations. A large fraction of protocols has converged in about 20 000 steps which is as little as 4 sweeps. The slowest protocol took 22 sweeps to end up in an optimum. A single sweep runs in just a few seconds on a laptop.}
\label{fig:Fidtrace}
\end{figure}

The efficiency of the method hinges completely upon the possibility to compute the local change in the fidelity (step 3) for all tweezer positions without having to recompute the entire time-evolution (step 2). The method can however not escape out of local optima in the fidelity landscape and we should run it for multiple initial random seeds. A typical set of optimization trajectories is show in Fig.~\ref{fig:Fidtrace}. While some initial seeds converge to a minimum very quickly others are substantially slower. Moreover, none of the final protocols are exactly the same but the corresponding fidelities are very similar. In fact all fidelities are somewhere  between $0.5311$ and $0.5364$, hence the best fidelity is only $1\%$ better than the worst. This behavior was also observed in~\cite{bukov_17} for controlling a single qubit. For a detailed discussion on the complexity and physics behind the optimization problem I refer to~\cite{bukov_17}.

\begin{figure}
\includegraphics[width=\columnwidth]{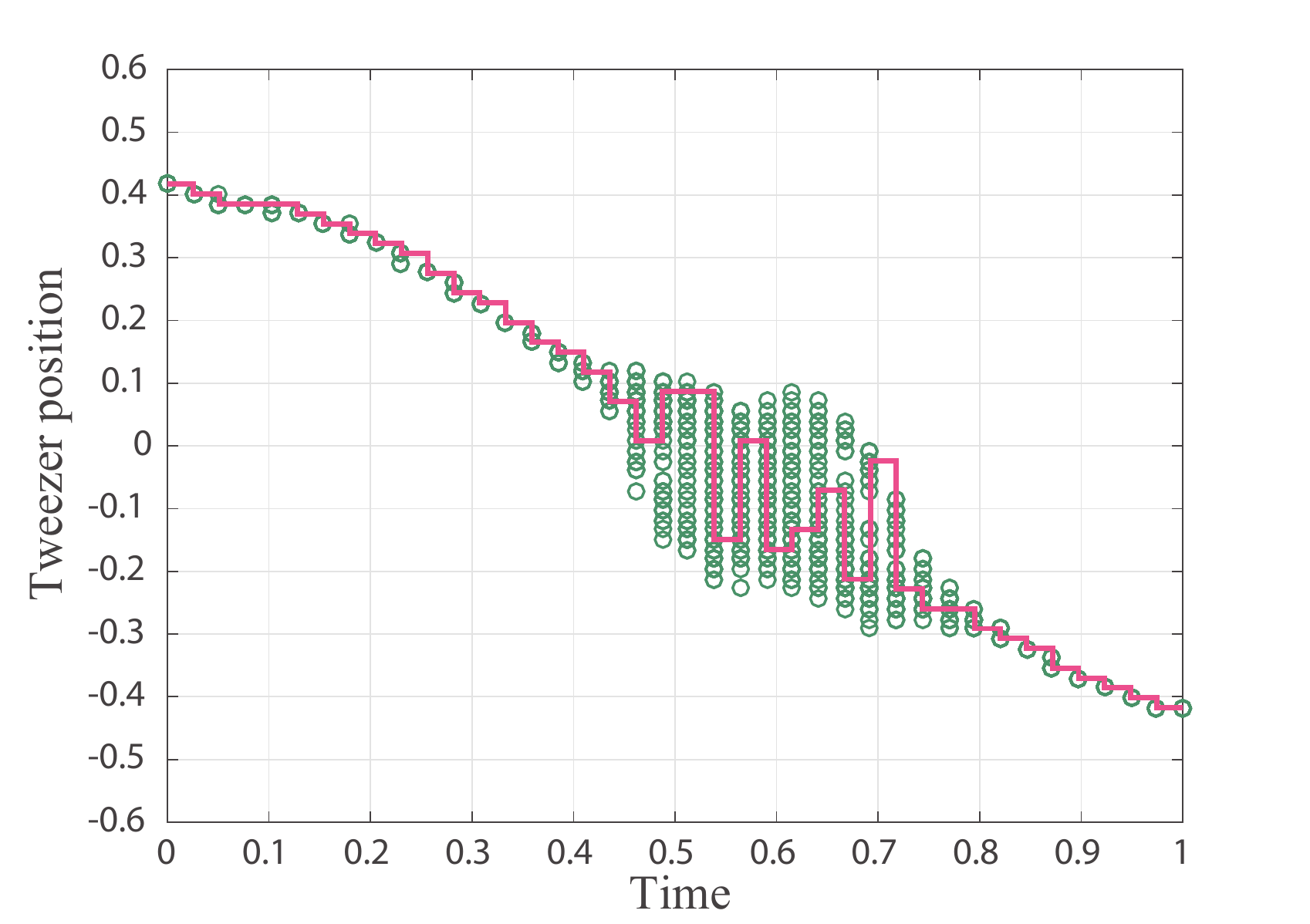}
\caption{{\bf Local optimal protocols for T=0.1} Figure shows 200 local optimal protocols that are found by stochastic optimization of a random initial seed. There are $M=128$ setting and $N=100$ timesteps. A representative protocol is shown by the pink line, the other protocols are scattered. }
\label{fig:protocols}
\end{figure}

For completeness, a scatter plot of protocols, obtained from random seeds, at $T=0.1$ is shown in Fig.~\ref{fig:protocols}. The beginning and the end of the protocol are always the same. Contrary to the counter-diabatic protocol, there is a region where the protocol actually oscillates (c.f. optimal results in Fig.~3(b) in Ref.~\cite{sorensen_16}). These oscillations are meant to remove the \emph{slushing} of the atom which still remained in the counter-diabtatic protocol. This can however be achieved in many ways with nearly the same fidelity, which explains the glasiness of the optimization landscape. 

\acknowledgements
I acknowledge financial support of the Research Foundation Flanders---FWO and the BU CMTV program. I thank F. Brosens and A. Polkovnikov for detailed remarks on the manuscript. The work benifited from many discussion with M. Bukov, A. Day, P. Mehta on machine learning. 

\bibliography{bibtweezer}

\end{document}